\documentclass{article}

\usepackage{PRIMEarxiv}

\usepackage[utf8]{inputenc} 
\usepackage[T1]{fontenc}    
\usepackage{hyperref}       
\usepackage{url}            
\usepackage{booktabs}       
\usepackage{amsfonts}       
\usepackage{nicefrac}       
\usepackage{microtype}      
\usepackage{lipsum}
\usepackage{fancyhdr}       
\usepackage{graphicx}       
\graphicspath{{media/}}     
\usepackage{subfigure}
\usepackage{multirow}
\usepackage{moreverb,url}

\title{A review of variable-pitch propellers and their control strategies in aerospace systems
}

\author{
Hanjie Jiang, Ye Zhou\thanks{Corresponding author} , Hann Woei Ho \\
School of Aerospace Engineering, Engineering Campus\\
Universiti Sains Malaysia, Pulau Pinang, Malaysia\\
\texttt{jianghanjie@student.usm.my}, \texttt{\{zhouye,aehannwoei\}@usm.my} 
}

\begin{document}
\maketitle

\begin{abstract}
The relentless pursuit of aircraft flight efficiency has thrust variable-pitch propeller technology into the forefront of aviation innovation. This technology, rooted in the ancient power unit of propellers, has found renewed significance, particularly in the realms of unmanned aerial vehicles and urban air mobility. This underscores the profound interplay between visionary aviation concepts and the enduring utility of propellers. Variable-pitch propellers are poised to be pivotal in shaping the future of human aviation, offering benefits such as extended endurance, enhanced maneuverability, improved fuel economy, and prolonged engine life. 
However, with additional capabilities come new technical challenges. The development of an online adaptive control of variable-pitch propellers that does not depend on an accurate dynamic model stands as a critical imperative. Therefore, a comprehensive review and forward-looking analysis of this technology is warranted. 
This paper introduces the development background of variable-pitch aviation propeller technology, encompassing diverse pitch angle adjustment schemes and their integration with various engine types. It places a central focus on the latest research frontiers and emerging directions in pitch control strategies. Lastly, it delves into the research domain of constant speed pitch control, articulating the three main challenges confronting this technology: inadequacies in system modeling, the intricacies of propeller-engine compatibility, and the impact of external, time-varying factors. 
By shedding light on these multifaceted aspects of variable-pitch propeller technology, this paper serves as a resource for aviation professionals and researchers navigating the intricate landscape of future aircraft development.
\end{abstract}


\section{Introduction}
Air propellers represent one of the oldest and enduring propulsion technologies in aviation history, with a rich and dynamic evolution. Since the pioneering days of aviation marked by the Wright brothers' historic Flyer 1 in 1903 \cite{wald2001wright}, the development of aeronautical propeller technology has remained closely intertwined with the progress of aircraft design. 
The 1930s witnessed a significant leap in both aircraft speed and engine power, driving the concurrent advancements in propeller technology \cite{wald2001wright, sand1973summary}. However, by the mid-1950s, the relentless development and refinement of turbojet engines, which do not rely on propellers to generate thrust, began to extend from military to civilian aviation. While this transition posed some challenges to propeller technology, their continued application in domains such as short take-off and landing (STOL) and long-endurance flight showcased their enduring relevance. The global oil crisis of the early 1970s rekindled interest in propeller-powered engines due to their energy efficiency \cite{sand1973summary}.

\begin{figure*}[htbp]
\centering
\subfigure
{
    \begin{minipage}[t]{.45\linewidth}
        \centering
        \includegraphics[scale=0.8]{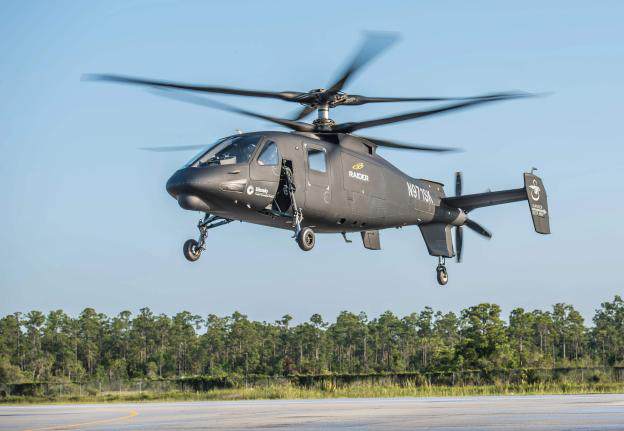} \\a) S-97 advanced high-speed helicopter (\cite{Lomawebsite}) 
    \end{minipage}
}
\subfigure
{
 	\begin{minipage}[t]{.45\linewidth}
        \centering
        \includegraphics[scale=0.8]{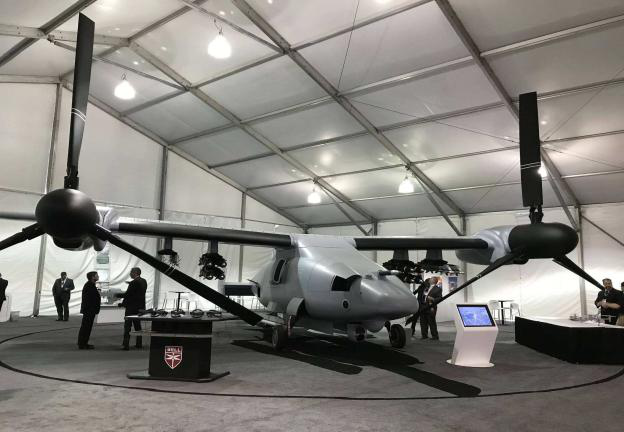} \\b) V-247 tilt-rotor UAV (\cite{Bellwebsite})
    \end{minipage}
}
\subfigure
{
 	\begin{minipage}[b]{.45\linewidth}
        \centering
        \includegraphics[scale=0.8]{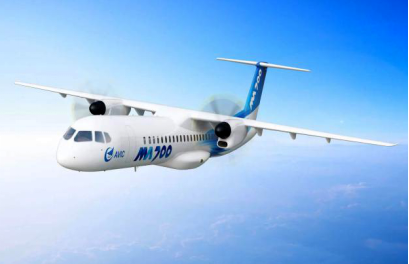} \\c) Ma700 advanced turboprop branch airliner (\cite{AVICwebsite})
    \end{minipage}
}
\subfigure
{
 	\begin{minipage}[b]{.45\linewidth}
        \centering
        \includegraphics[scale=0.8]{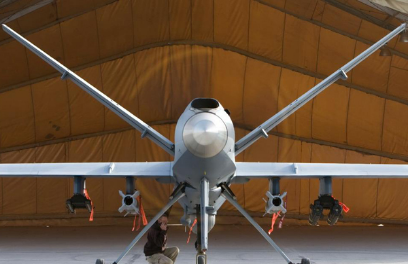} \\d) MQ-9 fixed wing military UAV (\cite{GAASwebsite})
    \end{minipage}
}
\subfigure
{
 	\begin{minipage}[b]{.45\linewidth}
        \centering
        \includegraphics[scale=0.0451]{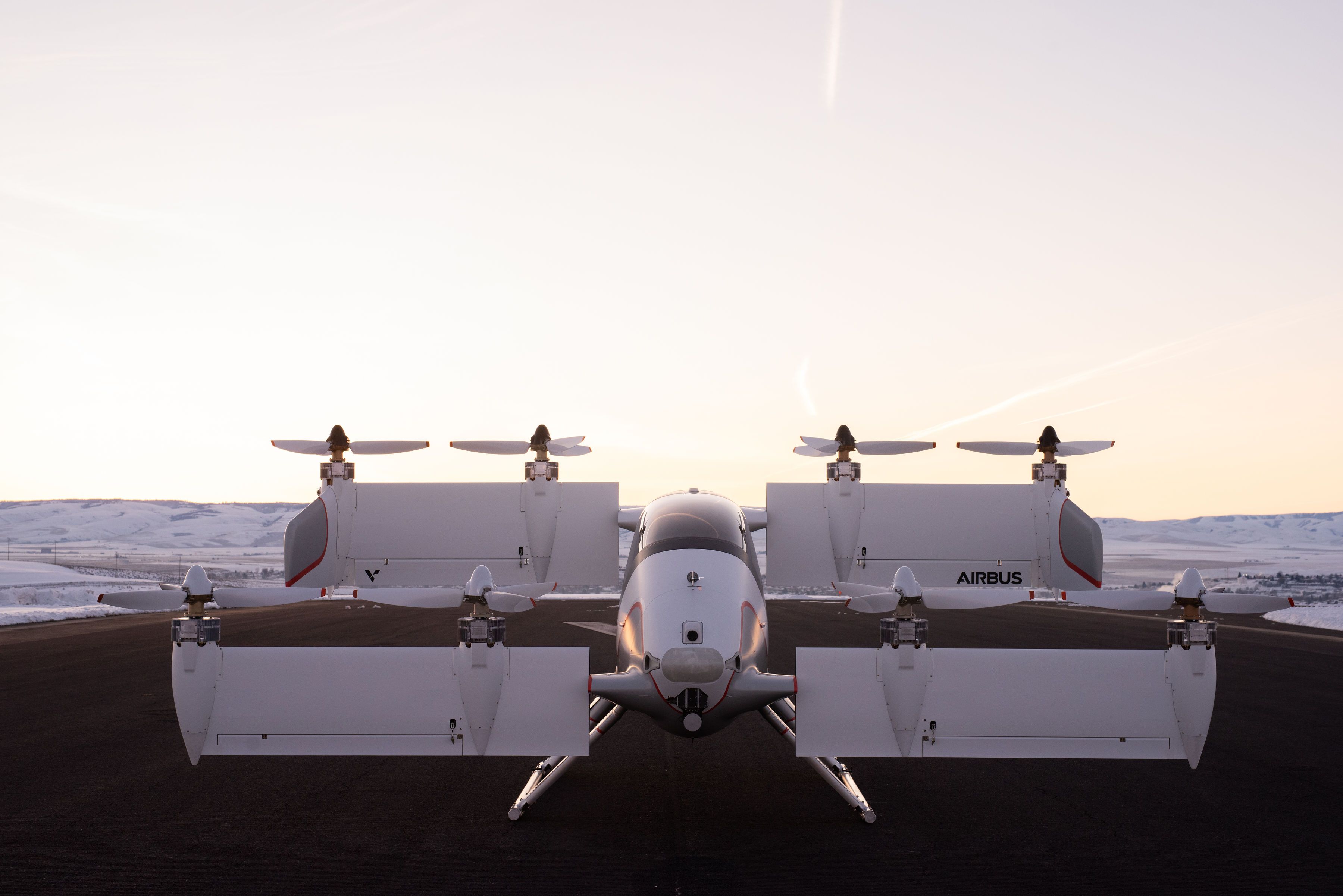} \\e) Airbus Vahana UAM Demonstrator (\cite{airbuswebsite})
    \end{minipage}
}
\subfigure
{
 	\begin{minipage}[b]{.45\linewidth}
        \centering
        \includegraphics[scale=0.8]{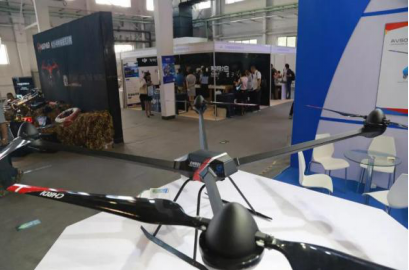}\\f) High mobility and altitude quadrotor (\cite{aerowebsite})
    \end{minipage}
}
\caption{Several types of aircraft using variable-pitch propellers.}
   \label{fig:1a}
\end{figure*}

Amidst the era of fixed-pitch propellers, the concept of variable-pitch propellers emerged and garnered extensive research attention. Unlike their fixed counterparts, which are optimized for a specific airspeed range around the design point, variable-pitch propellers offer the distinct advantage of adaptability to diverse flight conditions across the entire operational envelope. Consequently, aerospace engineers and researchers found variable-pitch propellers increasingly appealing and integrated them into various aircraft types \cite{Pistolesi1923Variable}. Notable examples include the groundbreaking S-97 advanced high-speed helicopter \cite{Lomawebsite} (Figure \ref{fig:1a}(a)), a collaborative effort by Sikorsky Helicopter and Boeing Defense. The coaxial main rotors of S-97 adopt rigid rotors,  which have no traditional flapping and lead-lag hinge, and retain the collective pitch control. Moreover, it incorporates a variable-pitch propulsion-propeller, enabling high-speed forward thrust as well as deceleration and even backward flight in level flight. Similarly, Bell's innovative V247 tilt-rotor Unmanned Aerial Vehicle (UAV) \cite{Bellwebsite} (Figure \ref{fig:1a}(b)) features a pair of variable-pitch propellers at the wingtips, facilitating vertical take-off and landing (VTOL), seamless transition, and efficient forward flight.

The cutting-edge Ma700 Advanced Turbo Propeller Branch Airliner, illustrated in Figure \ref{fig:1a}(c) \cite{AVICwebsite}, represents a remarkable achievement by the Aviation Industry Corporation of China. Furthermore, the MQ-9 UAV, developed by General Atomic Aviation Systems Corporation and showcased in Figure \ref{fig:1a}(d) \cite{GAASwebsite}, exemplifies a typical fixed-wing aircraft equipped with variable-pitch propellers. 
Variable-pitch propellers have also found their place in the rapidly evolving domain of electric vertical take-off and landing (eVTOL) aircraft \cite{2020An}, as evidenced by the Airbus Vahana Urban Air Mobility (UAM) depicted in Figure \ref{fig:1a}(e) \cite{airbuswebsite}. Currently, more than 100 technology startups worldwide are fervently engaged in developing UAM solutions, envisioning them as vital modes of future transportation, potentially supplanting traditional automobiles in certain scenarios. Notably, a majority of these UAM initiatives employ variable-pitch propellers to enhance performance and versatility. 
In addition to these larger aircraft, variable-pitch propellers have also made their mark in the realm of smaller aviation, including multi-rotor Unmanned Aerial Vehicles (UAVs), to improve their flight performances. Figure \ref{fig:1a}(f) showcases a high-mobility and high-altitude-capable multi-rotor aircraft developed by the China Helicopter Research and Development Institute (CHRDI) \cite{aerowebsite}, underscoring the adaptability of variable-pitch technology across a spectrum of aircraft types. 

The incorporation of variable-pitch propellers offers a host of advantages, including the expansion of flight envelopes, enhancement of flight performance, improved fuel efficiency, and extended engine life. However, fully harnessing these benefits necessitates addressing a range of intricate technical challenges. Beyond the study of propeller design and variable-pitch mechanisms, comprehensive research into the engines that drive these propellers and the intricate control systems uniting these three components is essential for maximizing their potential.

\section{Variable-pitch propellers}
\subsection{Basics of variable-pitch propellers} 

In 1872, Winham, a pioneer in British aviation, made a groundbreaking contribution by introducing variable-pitch propellers into fixed-wing aircraft, as documented in his annual report for the Aviation Society \cite{Pistolesi1923Variable}. However, it wasn't until the 1910s that people began to grasp the immense potential of variable-pitch propellers in enhancing engine power and efficiency. The progress of variable-pitch propellers, up to that point, had been largely constrained by the structural limitations of wooden propellers, remaining largely theoretical. 
The turning point came with the practical introduction of metal propellers in 1923, marking a significant milestone in aviation history. The demand for variable-pitch propellers surged, driven by the realization that traditional fixed-pitch propellers could not adequately meet the efficiency requirements across various flight phases \cite{2000Frank}.

\begin{figure}[h]
\centering                             
\includegraphics[width=0.9\linewidth]{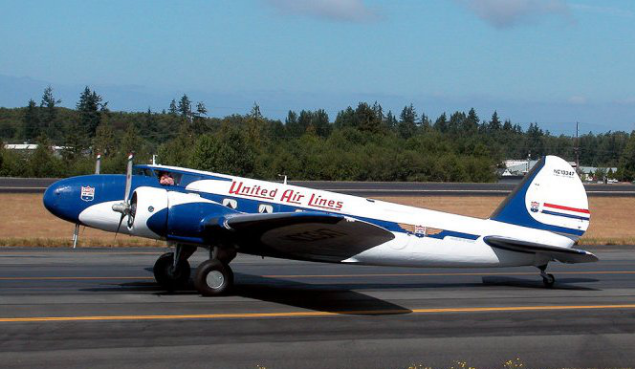} 
\caption{Boeing 247 airliner \cite{van2011boeing}}
\label{fig:2}   
\end{figure}   

A few years later, the US Air Force embarked on a mission to optimize propeller efficiency throughout the entire flight envelope by providing a befitting blade angle for each section of the flight envelope. This endeavor aimed to surpass the performance of fixed-pitch propellers. It became evident that maintaining precise control over engine speed at different flight speeds held the key to enhancing both propeller and engine efficiency. Equally crucial was the ability to regulate engine speed to ensure consistent power output, spanning from takeoff to maximum flight speed.

In a pivotal development in 1933, the automatic variable-pitch propeller made its debut in the Boeing 247, depicted in Figure \ref{fig:2}. This innovation brought about substantial improvements in performance and adaptability. Implementing a variable-pitch mechanism during takeoff and cruise flight yielded remarkable benefits, including a 20\% reduction in takeoff run distance, a 22\% increase in climb rate, a 5\% boost in cruise speed, and an impressive 1220-meter gain in ceiling altitude \cite{van2011boeing,2000Frank}. This technology rapidly evolved into a sophisticated constant-speed propeller system, seamlessly adjusting the pitch in real-time to match flight speed variations.

Variable-pitch propellers employ a mechanism that synchronously rotates their blades around each blade handle's axis. In the realm of aircraft propulsion, these variable-pitch propellers can be categorized into two fundamental types based on their blade rotation mechanisms:

\begin{enumerate}
\item[(i)] Hydraulic variable-pitch propellers: These systems utilize hydraulic pressure to drive the blade rotation mechanism and are predominantly employed in larger and medium-sized aircraft. 
\item[(ii)] Electric variable-pitch propellers: These systems rely on electric motors to drive the blade rotation mechanism and find frequent use in smaller aircraft, especially UAVs.
\end{enumerate}

\subsubsection{Hydraulic variable-pitch propellers}

\begin{figure}[h]
\centering                            
\includegraphics[width=0.9\linewidth]{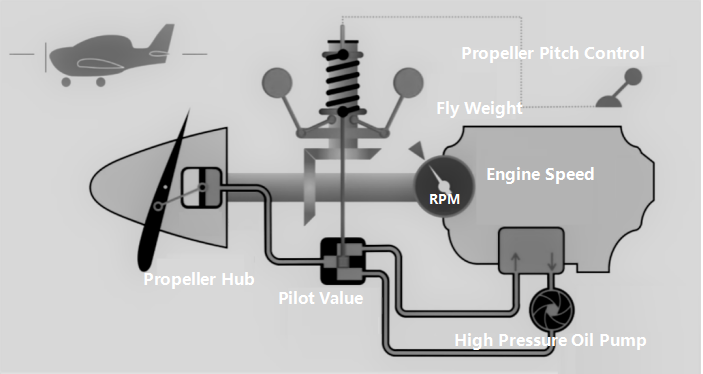} 
\caption{Hydraulic pitch control system \cite{Fclubwebsite}.}
\label{fig:3}   
\end{figure}

Hydraulic pitch propellers employ hydraulically operated governors to effect changes in blade angles. These governors incorporate rotating flyweights, a preloaded spring, and a driving gear, collectively constituting a critical component known as the Constant Speed Unit (CSU) \cite{Fclubwebsite}. The engine crankshaft drives the governor, causing the flyweights to respond to alterations in engine speed. When Revolutions Per Minute (RPM) of the engine increases, centrifugal forces compel the flyweights to move outward, and vice versa, as illustrated in Figure \ref{fig:3}. Pilots exercise control over the extent of flyweight movement through a lever known as the propeller pitch control. Adjusting this lever tensions the spring connected to the flyweights. The constant speed device ensures that the propeller maintains a constant speed during flight. The governor's flyweights are intricately linked to the pilot valve, which governs the flow of oil to or from the propeller hub, thus determining the necessary propeller blade angle.

When the propeller pitch control is adjusted in a rearward direction, the pilot's intention is to decrease the target rotational speed of the propeller. In response, the flyweights move outward, causing the pilot valve to open, and permitting the flow of oil into the propeller hub.  Increasing blade angles will increase the air pressure on the propeller, which in turn demands more engine torque. As a result, the RPM decreases, and the flyweights return to their equilibrium position. Once the pilot valve is closed, both the vane angle and engine speed stop changing. 
Conversely, when the propeller pitch control is shifted forward, the pilot intends to increase the target rotational speed. In this scenario, the flyweights move inward, facilitating the outflow of oil from the propeller hub. Consequently, the blade angle decreases, causing the propeller to take a smaller ``bite" of the air and necessitating less engine torque. As a result, the RPM increases, and the flyweights return to their equilibrium position. Once again, the pilot valve is closed, and both the blade angle and engine RPM are held steady \cite{Fclubwebsite, muhlbauer2013hydraulically, heinrich1950hydraulically}.

\subsubsection{Electric variable-pitch propeller}

\begin{figure}[h]
\centering                                 
\includegraphics[width=\linewidth]{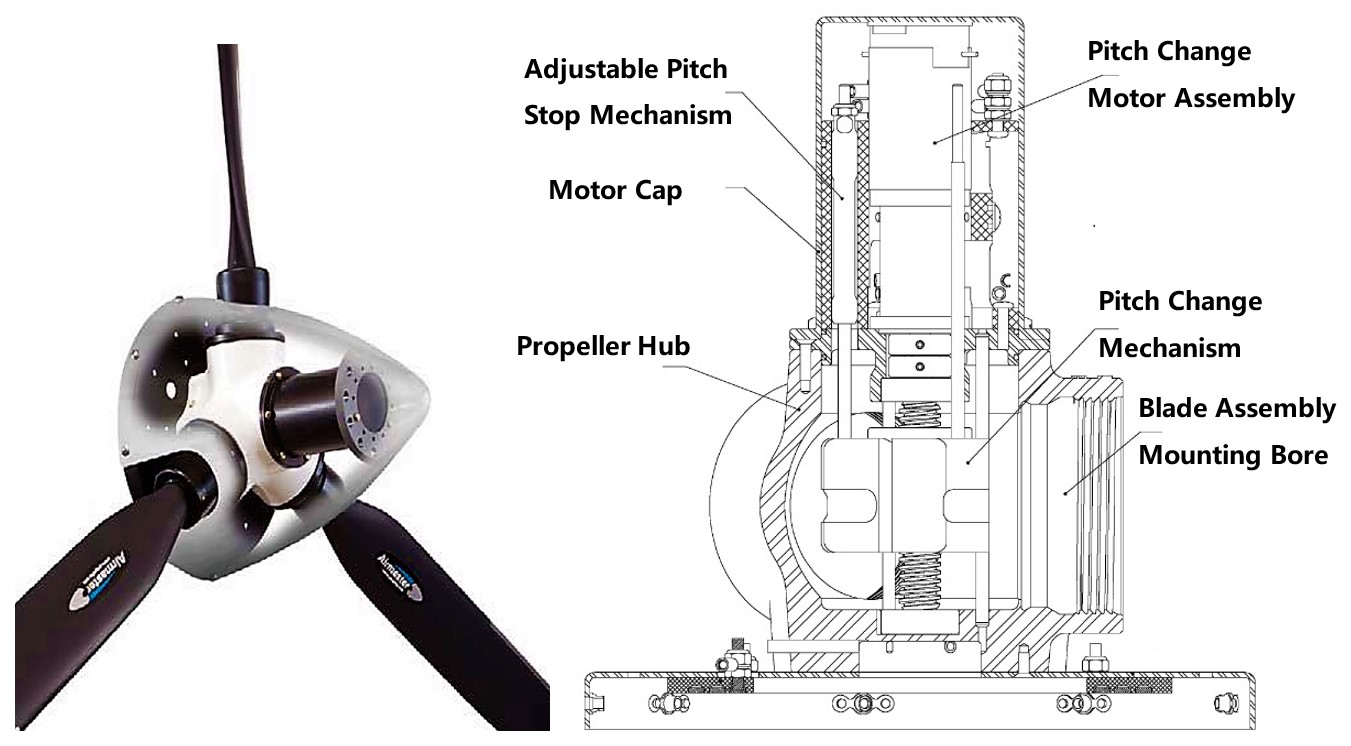} 
\caption{Electric pitch control system \cite{OPERATORMANUAL}.}
\label{fig:4}   
\end{figure}

The electrically operated pitch change mechanism system incorporates a control system that passes electric power to the propeller via a sensor/brush assembly and a specially constructed slip-ring assembly. The assemblies are mounted separately on the aircraft engine and the spinner back-plate \cite{OPERATORMANUAL}, as illustrated in Figure \ref{fig:4}.  The system's pitch adjustment mechanism is orchestrated by an electric servo motor working in tandem with a planetary gearbox. 
The pitch variation mechanism comprises a precision control screw mechanism responsible for regulating the position of the pitch variation slider. This slider, positioned along the axis of the propeller hub, in turn, impels the cams affixed to the assembly base of each propeller blade. As a result, the pitch mechanism orchestrates precise adjustments to the angle of each blade.

Similar to the hydraulic pitch mechanism, the electric counterpart also features a constant speed controller, harnessed through a solid-state microprocessor. This governor orchestrates adjustments to the blade or pitch angle, ensuring a consistent engine and propeller speed. Propeller speed is meticulously gauged through a solid-state magnetic sensor mounted on the aircraft engine \cite{OPERATORMANUAL,1960Variable}. 
The electronic governor operates by comparing the current propeller speed with the speed specified by the pilot, employing a control loop to minimize any disparities. The nonlinear response of this control loop is finely tuned through a range of parameters designed in the controller software. These control parameters dictate how the controller responds to variations in speed error, affecting factors such as the error's magnitude and whether it is diminishing or increasing.

\subsection{Engines with variable-pitch propellers}

The aviation industry currently encompasses more than ten types of aero-engines, as illustrated in Table \ref{tab:11}. These engines are categorized based on different thrust generation methodologies and engine structures. Irrespective of their specific design, all aero-engines achieve propulsion by exerting a propulsive force in opposition to a particular composition of fluid, commonly referred to as the propulsive working fluid \cite{Rolls1996THE,Van2004Internal}. 
One classification involves the direct ejection of the fluid passing through the engine, such as turbojet engines, rocket engines, and similar variants. In contrast, the second major category entails engines that drive propellers or other power devices to transmit momentum into the surrounding ambient air. This category encompasses engines like turboshaft engines and piston engines, among others.

\begin{table*}[htbp]
\caption{Aircraft Engine Types} 
\centering
\resizebox{1\linewidth}{!}{
\begin{tabular}{c c c c c}
\hline
\multicolumn{2}{c}{Engine Type}&Means of Compression&Engine Working Fluid&Propulsive Working Fluid\\  
\hline
\multirow{5}*{Gas Turbine Engine}&Turbojet Engine&Turbine-driven Compressor&Fuel/Air Mixture&Fuel/Air Mixture\\
\cline{2-5}
&\textbf{\textit{Turboprop Engine}}&\textbf{\textit{Turbine-driven Compressor}}&\textbf{\textit{Fuel/Air Mixture}}&\textbf{\textit{Ambient Air}}\\
\cline{2-5}
&Turbofan Engine&Turbine-driven Compressor+Fan&Fuel/Air Mixture&Fuel/Air Mixture\\
\cline{2-5}
&\textbf{\textit{Turboshaft Engine}}&\textbf{\textit{Turbine-driven Compressor}}&\textbf{\textit{Fuel/Air Mixture}}&\textbf{\textit{Ambient Air}}\\
\cline{2-5}
&Propfan Engine&Turbine-driven Compressor+Fan&Fuel/Air Mixture&Fuel/Air Mixture
+Ambient Air\\
\hline
\multicolumn{2}{c}{Ramjet Engine}&Ram Compression&Fuel/Air Mixture&Fuel/Air Mixture\\  
\hline
\multicolumn{2}{c}{Pulse-jet Engine}&Compression Due to Combustion&Fuel/Air Mixture&Fuel/Air Mixture\\  
\hline
\multicolumn{2}{c}{\textbf{\textit{Wankel Engine}}}&\textbf{\textit{Rotation of Rotors}}&\textbf{\textit{Fuel/Air Mixture}}&\textbf{\textit{Ambient Air}}\\  
\hline
\multicolumn{2}{c}{\textbf{\textit{Piston Engine}}}&\textbf{\textit{Reciprocating Action of Pistons}}&\textbf{\textit{Fuel/Air Mixture}}&\textbf{\textit{Ambient Air}}\\  
\hline
\multicolumn{2}{c}{\textbf{\textit{DC Motor}}}&--&--&\textbf{\textit{Ambient Air}}\\  
\hline
\end{tabular}}
\label{tab:11}
\end{table*}

Beyond the generation of thrust, aircraft engines require a source of energy to sustain their operation. In the case of aero-engines, nearly all of them rely on the expansion of fluids created through the combustion of fuel in the presence of an oxidizer or air. This resulting composition, referred to as the engine working fluid, serves as the vital energy source. Various types of aviation engines employ diverse methods of compressing air to facilitate its entry into the engine. 

Several engine types, as indicated in Table 1, are closely integrated with propellers, including those equipped with variable-pitch propellers. These engine categories encompass turboprop engines, turboshaft engines, piston engines, Wankle engines, and DC motors \cite{Rolls1996THE,Van2004Internal}. These aero-engines act by driving propellers to expel ambient air in a manner that generates forward propulsion or lift to counteract the force of gravity.

It's worth noting that piston engines and wankle engines adopt a similar operational principle, driven by the periodic movement of rotors or pistons. While the DC motor doesn't align with the conventional concept of engines, its significance cannot be understated, particularly in the context of its indispensable role in the advancement of fields such as multi-electric aircraft, UAVs, eVTOLs, and more.

\begin{figure}[h]
\centering                             
\includegraphics[width=0.8\linewidth]{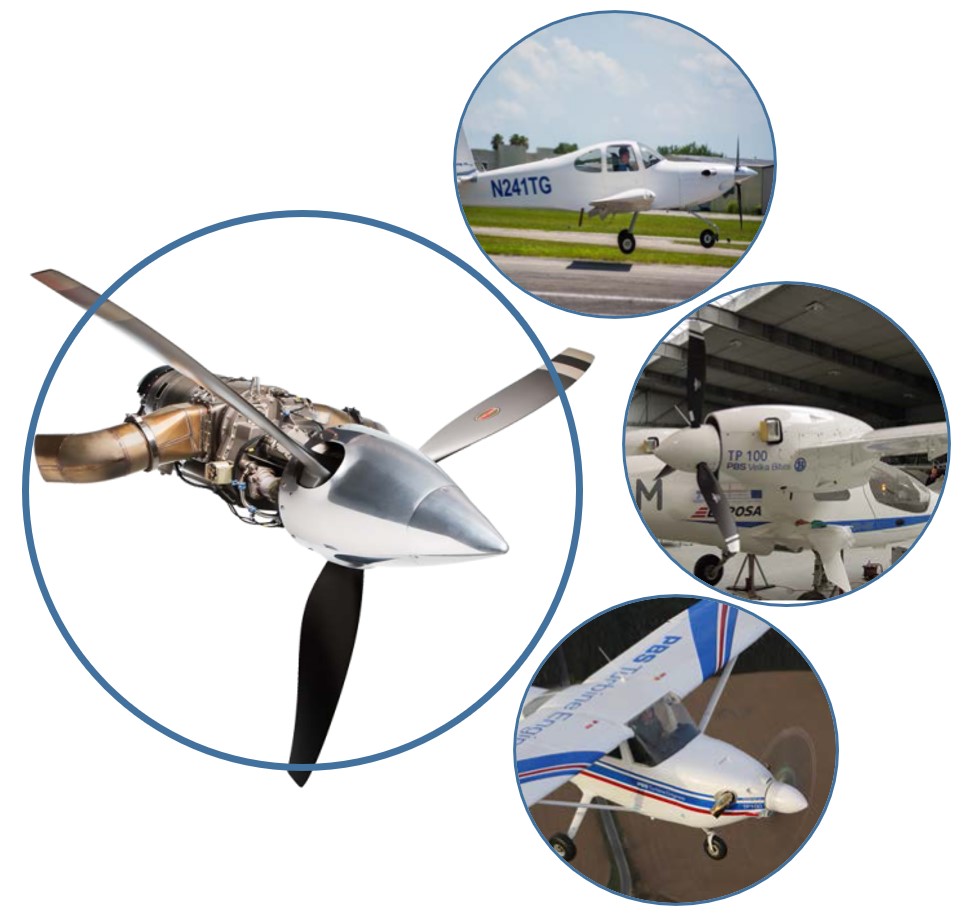} 
\caption{Turboprop engine for test aircraft, ultralight aircraft and UAVs \cite{pbstp}.}\label{fig:5}   
\end{figure}  

\subsubsection{Turboprop Engine}
Turboprop engines represent a subtype of gas turbine engines extensively utilized in aircraft of diverse sizes. These engines primarily allocate a significant portion of their power output to drive an external variable-pitch propeller. Typically, turboprop engines find favor among smaller aircraft categories, including test aircraft, ultralight aircraft, and UAVs, as depicted in Figure \ref{fig:5}.

The anatomy of a turboprop engine encompasses key components such as an intake, a compressor, a combustion chamber, a turbine, and a nozzle. The airflow is channeled through the intake port and flows into the compressor \cite{Rolls1996THE}. Fuel is added to the compressed air within the combustion chamber, igniting the working fluid that proceeds into the turbine. Here, the kinetic energy generated serves a dual purpose: it propels the propeller while concurrently sustaining the compressor's operation by supplying air to the combustion chamber. The remaining energy cgenerates a small amount of thrust outside the propeller via the nozzle. Most turboprop engines feature a hydraulic variable-pitch propeller paired with a constant-speed mechanism.

\subsubsection{Turboshaft Engine}

\begin{figure}[h]
\centering                                 
\includegraphics[width=0.8\linewidth]{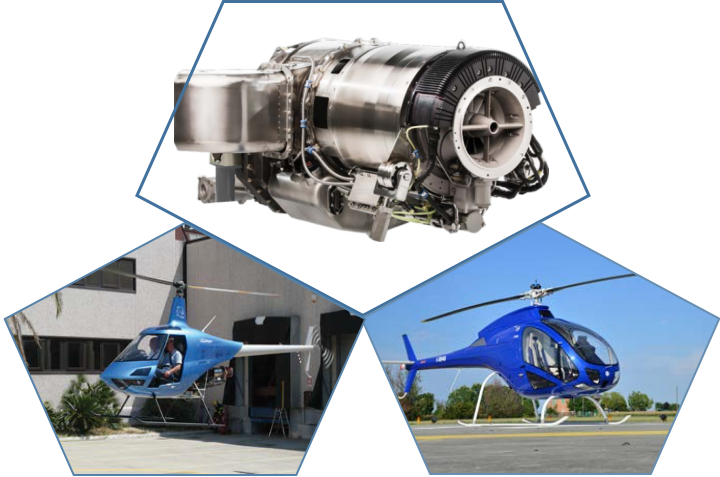} 
\caption{Turboshaft engine designed for light helicopters and UAVs \cite{pbsts}.}\label{fig:6}   
\end{figure}

Turboshaft engines, another category of gas turbine engines, are optimized for maximizing shaft power output. These engines find extensive use in helicopters and rotor-type aircraft relying on a shaft drive, as exemplified in Figure \ref{fig:6}. Turboshaft engines hold a distinct advantage by delivering impressive power-to-weight ratios exceeding 2.5 kW/kg \cite{johnson1991advanced, Rolls1996THE}. In terms of raw power production, turboshaft engines can reach substantial magnitudes. Presently, these engines can generate up to 6,000 or even 10,000 horsepower, a stark contrast to the relatively lower power output of piston engines. Economically, while turboshaft engines consume slightly more fuel compared to top-performing piston engines, this is partly compensated by the fact that jet fuel is cheaper than petrol. Nevertheless, it is essential to acknowledge that the manufacturing of turboshaft engines is intricate and costly, representing a significant drawback in their widespread adoption.

The turboshaft engine shares fundamental functions and structural similarities with other turbogenerator engines. In terms of its structural composition, the turboshaft engine adheres to the fundamental framework of a gas generator, featuring components such as an inlet, compressor, combustion chamber, and exhaust nozzle. However, a defining characteristic of the turboshaft engine is the incorporation of a free turbine. This unique turbine does not drive the compressor; instead, it serves the primary purpose of power generation.

The alteration in rotor speed within the engine induces significant variations in centrifugal forces. To ensure stable operation, the rotor needs to be designed to maintain a consistent speed, driven by the free turbine. Any fluctuations in power output are finely tuned by adjusting the blade pitch accordingly, allowing for precise control of the engine's performance.

\subsubsection{Piston engine}

The piston engine stands is the most common type of power plant found in today's world, from automobiles and boats to an array of self-powered machinery. Regardless of whether it's fueled by gasoline, alcohol, or diesel, the piston engine is the driving force behind many of mankind's prized mechanical possessions. 
This engine design employs pistons connected to a crankshaft through connecting rods, enabling the transmission of power. Fuel and air are drawn into the engine through the carburetor or fuel injection system, eventually entering the combustion chamber. Here, the mixture is ignited by spark plugs, initiating a downward motion of the piston. This, in turn, sets the crankshaft into motion, producing the mechanical power that drives the machinery.

\begin{figure}[h]
\centering                                 
\includegraphics[width=\linewidth]{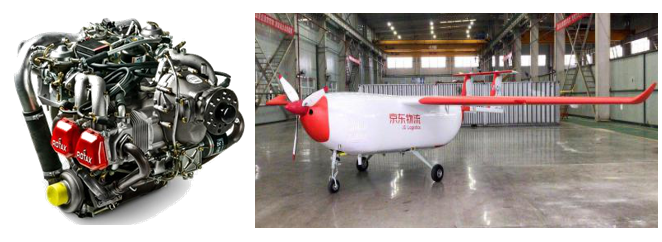} 
\caption{The JD Logistics UAV and its piston engine \cite{JingDong}.}\label{fig:7}   
\end{figure}

In the mid-1940s, piston aero-engines began to yield ground to gas turbine engines in military aircraft and large-scale civil aircraft \cite{Van2004Internal}. Nevertheless, piston aero-engines continue to find widespread use in light, low-speed aircraft, helicopters, and UAVs, primarily due to their exceptional cost-effectiveness. Figure \ref{fig:7} showcases the logistics UAV deployed by Jingdong, which employs a Rotax piston engine to reduce manufacturing and operational costs.

\subsubsection{Wankel engine}

The Wankel engine, also known as the rotary engine, owes its invention to the German engineer Felix Wankel (1902-1988). Drawing on insights from prior research, he successfully tackled critical technical challenges and successfully developed the first functional Wankel engine \cite{1971The}. What sets the Wankel engine apart is its unique reliance on the rotational movement of a triangular rotor to govern the processes of compression and exhaust, a stark departure from the linear motion characteristic of traditional reciprocating piston engines. 
In this ingenious design, as the center of the triangular rotor orbits around the center of the output shaft, the rotor itself undergoes rotation around its own center. This triangular rotor effectively partitions the rotor housing into three distinct chambers, each sequentially carrying out intake, compression, power generation, and exhaust functions. Remarkably, the triangular rotor completes three cycles during a single revolution.

\begin{figure}[h]
\centering                                 
\includegraphics[width=0.9\linewidth]{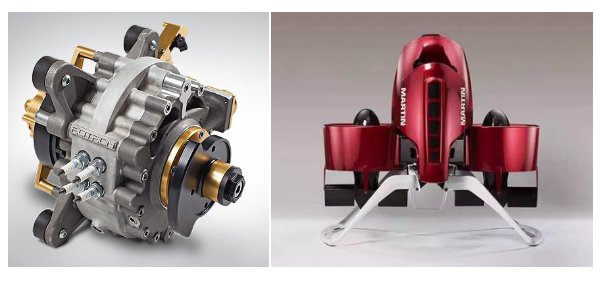} 
\caption{The Martin Jetpack and its wankel engine \cite{mtjp,rotron}.}\label{fig:8}
\end{figure}

Compared to conventional four-stroke engines, which perform work only once every two revolutions, the Wankel engine boasts a distinct advantage with its high horsepower-to-volume ratio. Furthermore, due to the rotational running characteristics of the Wankel engine, it doesn't necessitate precise crankshaft balancing to achieve higher operational speeds. This engine features a minimal number of moving parts, just two in total, as opposed to the more than 20 moving parts found in typical four-stroke engines, including components like intake and exhaust valves. This streamlined structure not only simplifies the engine but also significantly reduces the likelihood of failure.

While the use of Wankel engines in the automotive industry has been restricted due to their relatively high specific fuel consumption, susceptibility to wear, and elevated cost, they have found a niche in the aviation field. Notably, their small size, light weight, and favorable vibration characteristics make them well-suited for specialized aviation applications. A typical example is the Martin Jetpack, which employs a Wankel engine to fulfill the size and power-to-weight ratio requirements of its power plant, as depicted in Figure \ref{fig:8}.

\subsubsection{DC motor}

DC motors harness the principles of electricity and magnetic fields to generate shaft power\cite{rainbow,xduam,wong2021design}. In its simplest configuration, this motor relies on a pair of magnets with opposing polarities and a coiled wire acting as an electromagnet. The interplay of attraction and repulsion between these magnets furnishes the torque necessary to set the motor in motion.

\begin{figure}[h]
\centering                                 
\includegraphics[width=\linewidth]{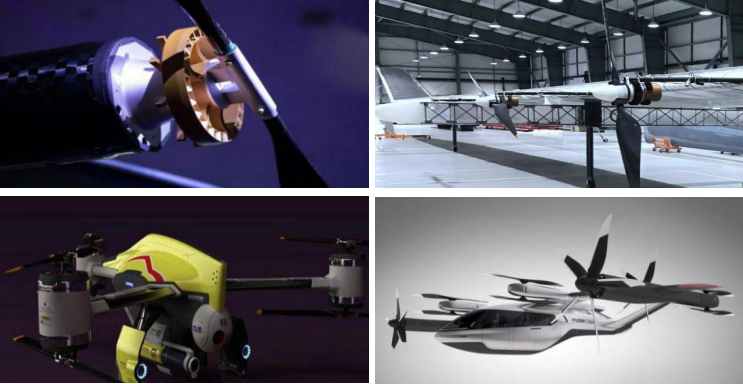} 
\caption{DC motors are used in the field of UAVs and UAMs \cite{rainbow,xduam}.}\label{fig:9}
\end{figure}

Notably, brushless DC motors have become the powerhouses driving a myriad of UAV propulsion systems, particularly in the realm of multi-rotor UAVs and various clean energy UAV setups. The utilization of alternative energy sources such as solar power, hydrogen energy, and fuel cell technology in the aviation sector hinges on the performance of DC motors. Additionally, the influence of DC motor power systems extends to the evolution of general aviation, encompassing domains like ultralight aircraft and eVTOL aircraft, as depicted in Figure \ref{fig:9}.

Brushless motors stand out due to their commendable dynamic response, impressive power-to-weight ratio, and environmentally friendly attributes when contrasted with traditional fuel engines. Presently, the development and application of brushless motors are predominantly influenced by advancements in battery technology. In specific scenarios demanding extended endurance, piston engines continue to hold sway, with certain applications opting for a hybrid configuration that combines piston engines with DC motors.

\section{Variable-pitch control and engine control}

\subsection{Variable-pitch control} 

The linchpin in the seamless integration of variable-pitch propellers with various engine types resides in precise variable-pitch control. Numerous research endeavors have contributed a diverse array of methods for controlling variable-pitch propellers \cite{Day1990Aircraft,French1945Engine,Pedrami2018SYSTEM}. Among these, the pitch angle controller, founded on Proportional-Integral (PI), Proportional-Differential (PD), and Proportional-Integral-Differential (PID) strategies, enjoys widespread application \cite{2015Advanced,2017Pitch,2018Systematic,2013Dynamic}. Presently, the conventional PID control method is the preferred choice in pitch control systems, well-suited to specific operational conditions. However, its adaptability falters when confronted with changing operating conditions, rendering controller parameter adjustments a challenging endeavor \cite{622006Character}.

For enhanced robustness in addressing nonlinear challenges, Linear Quadratic Gaussian (LQG) and Sliding Mode Control (SMC) techniques have been harnessed for pitch angle control \cite{2010Sliding}. SMC, in particular, stands out as an effective approach for designing robust control methodologies tailored to the intricacies of complex variable-pitch nonlinear systems. It endeavors to resolve fundamental issues such as time delays, parameter uncertainties, and disturbances \cite{602020Multiloop}. However, SMC often has converging problems when approaching the sliding surfaces and relies on fixed control laws that are not easily adaptable to changing system dynamics or varying operational conditions. This is a similar problem in aircraft control. These limitations have spurred the exploration of adaptive control methodologies, such as Incremental Nonlinear Dynamic Inversion (INDI) \cite{zhou2021extended} and Adaptive Model Predictive Control (AMPC) \cite{jiang2023adaptive}, which exhibit the capacity to adapt and fine-tune control strategies in response to evolving system behaviors, external disturbances, and uncertainties. This adaptability positions INDI and AMPC as promising candidates for further enhancing the precision and versatility of variable-pitch control systems.

Intelligent control techniques have also found a niche in modeling and controlling variable-pitch propellers, particularly in the context of nonlinear dynamic systems. Neural network and fuzzy logic methods have emerged as potent tools when grappling with highly nonlinear systems. Artificial Neural Networks (ANN) exhibit remarkable precision in nonlinear control under specific system conditions, relying on trainable information \cite{Kasabov2002Foundations,2010A}. Building on this foundation, machine learning algorithms, including Reinforcement Learning (RL), have been applied to pitch control challenges, bolstering the controller's online learning capabilities, reducing model requirements, and delivering commendable control precision \cite{2020Reinforcement,37Ng2003Autonomous}. 
Recent research efforts have witnessed a surge in RL studies, aiming to enhance learning efficiency and adaptability in the face of nonlinearities and uncertainties \cite{zhou2018online,zhou2017launch}. These advancements have positioned RL as a promising solution, particularly when confronted with the intricacies of optimal adaptive control problems inherent in variable-pitch control systems.

\subsection{Control strategies for aircraft with variable-pitch propellers}

Beyond the fundamental propeller pitch control methods, the realm of aircraft equipped with variable-pitch propellers has witnessed a surge in enthusiasm for innovative control strategies. These aircraft, including helicopters, rotors, and fixed-wing platforms, harness variable pitch technology and appropriate control strategies in pursuit of specific operational advantages \cite{562016Research,572019Propeller,582018Nonlinear,592015A,612014Variable}. 
A noteworthy development in this field involves the development of practical online optimization algorithms aimed at minimizing power consumption within the propulsion system across a range of thrust settings \cite{382013Online}. This pioneering approach has resulted in a remarkable 25\% enhancement in power efficiency and was effectively applied to a variable-pitch propeller actuated by a DC motor.

Furthermore, the groundwork for the Variable-Pitch Propeller Drive Controller (VPPDC) has been laid \cite{392019Controller}, extending the standard drive configuration to enhance energy efficiency and prolong flight duration. VPPDC leverages precise knowledge of propeller and brushless motor characteristics, employing an online optimization algorithm to calculate the optimal blade angle. This calculated angle minimizes power consumption within the electric propulsion system for the given thrust value, offering an innovative solution for improved flight efficiency.

In addition, researchers have delved into the domain of flight dynamics modeling and controller design for variable-pitch Quadrotors \cite{402016Flight}. Their approach incorporated three control loops, with an additional loop addressing the challenges of control allocation stemming from the non-trivial relationship between variable-pitch and rotor forces. This multi-loop controller framework enhances the maneuverability and control precision of variable-pitch Quadrotors, further expanding the application possibilities of this technology.

Despite the advancements in variable-pitch flight control strategies, there are still several promising control methods used in flight control that have yet to be extensively applied in the realm of variable-pitch aircraft. One such method is INDI \cite{zhou2021extended, HoRSS23}, which excels in handling unknown dynamics, changing conditions, and disturbances by iteratively updating control laws. When used in complex dynamic systems like aircraft with variable-pitch propellers, INDI with a cascaded control structure can be employed to handle the intricacies of these systems while providing effective control and adaptation. 
Another intriguing approach is Adaptive Critic Designs (ACDs) \cite{zhou2022efficient}, which combines RL and neural networks to adapt and optimize control policies based on the aircraft's performance and environmental factors. 
Additionally, Hierarchical Reinforcement Learning (HRL) offers the potential to create complex control hierarchies that manage various aspects of flight simultaneously \cite{zhou2022online}, providing a comprehensive approach to aircraft control. These control methods have shown promise in addressing complex and dynamic flight scenarios and may hold the key to further enhancing the capabilities of variable-pitch aircraft.

\subsection{Benefits to engineering applications}

Variable-pitch propeller control has garnered increasing attention within the aerospace engineering domain, driven by its potential to deliver substantial benefits. These anticipated advantages for the aerospace sector can be succinctly outlined as follows:

\begin{enumerate}
\item[(i)] Improve the endurance performance of the aircraft
Findings from experimental research conducted at the College of Automation Engineering, Nanjing University of Aeronautics and Astronautics \cite{412016Control} have illustrated that variable-pitch quadrotors can simultaneously improve endurance performance and positioning accuracy when compared to fixed-pitch counterparts by using steady-state identification method with minimum power consumption, as shown in Figure \ref{fig:10}. 
Furthermore, the VPPDC research at Ostrava University of Technology has showcased the significant impact of variable-pitch propulsion units (VPPU) equipped with integrated algorithms, such as the Adaptive Pitch Control Algorithm (APCA) or Pitch Control Algorithm (PCA), in augmenting hover and flight times \cite{392019Controller}.

\begin{figure}[h]
\centering                                 
\includegraphics[width=0.8\linewidth]{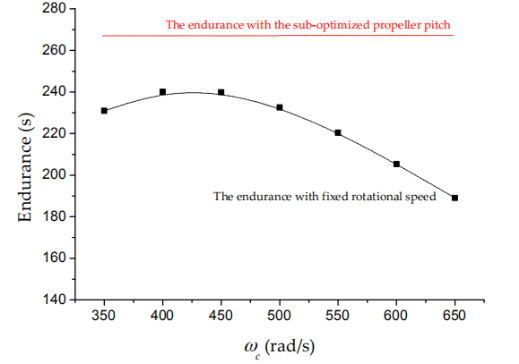} 
\caption{The endurance performance versus propeller pitch angles \cite{412016Control}.}
\label{fig:10}
\end{figure}

\item[(ii)] Enhance the aircraft maneuverability

A series of studies conducted by Cutler and his team at Massachusetts Institute of Technology (MIT) have shed light on the notable dynamic distinctions in thrust output between fixed-pitch and variable-pitch propellers \cite{422011Comparison,432012Actuator,442012designandcontrol}. Variable-pitch actuation offers substantial advantages over fixed-pitch quadrotors, particularly in terms of enhancing thrust response and enabling swift and efficient force reversal. 
Variable-pitch propellers confer a pivotal advantage to quadrotors—the capacity to generate reverse thrust. This capability empowers the vehicle to execute maneuvers like flying upside down and enables rapid deceleration by momentarily reversing the propeller pitch. In variable-pitch mode, the quadrotor exhibits exceptional tracking precision, with just 1\% overshoot compared to the substantial 60\% overshoot observed in fixed-pitch mode \cite{442012designandcontrol}. The improved tracking performance is attributed to the ability to achieve significant negative accelerations through pitch control.

Furthermore, the Indian Institute of Technology has developed a hybrid UAV combining a variable-pitch quadrotor and fixed-wing configuration for express transportation \cite{2018Systematic}. Variable-pitch control technology bestows this platform with exceptional maneuverability in logistics applications, enhancing its capability to navigate and adapt to diverse scenarios effectively.

\item[(iii)] Improve the fuel economy of the power system

The external characteristics of the 492Q piston engine, as depicted in Figure \ref{fig:11}, illustrate the power, torque, and power-specific fuel consumption curves of the engine under full load conditions (with the gasoline engine at full throttle) as functions of speed \cite{452000simulated}. The symbols $\mathnormal{g_c}$ and $\mathnormal{P_c}$ denote the engine-specific fuel consumption and shaft power, respectively, while $\mathnormal{T_c}$ represents the shaft torque of the piston engine. These parameters exhibit variations across engine speeds, spanning from 1000 RPM to approximately 3700 RPM. 
Notably, at around 3000 RPM, the engine exhibits nearly the minimum power-specific fuel consumption, accompanied by significant effective power and appropriate torque output. These data represent typical piston engine performance characteristics. In practice, engines should not operate at their maximum speeds for extended durations. For the 492Q engine, maintaining a constant speed of around 3000 RPM allows for sufficient power and torque output while effectively reducing fuel consumption.

\begin{figure}[h]
\centering                            
\includegraphics[width=0.75\linewidth]{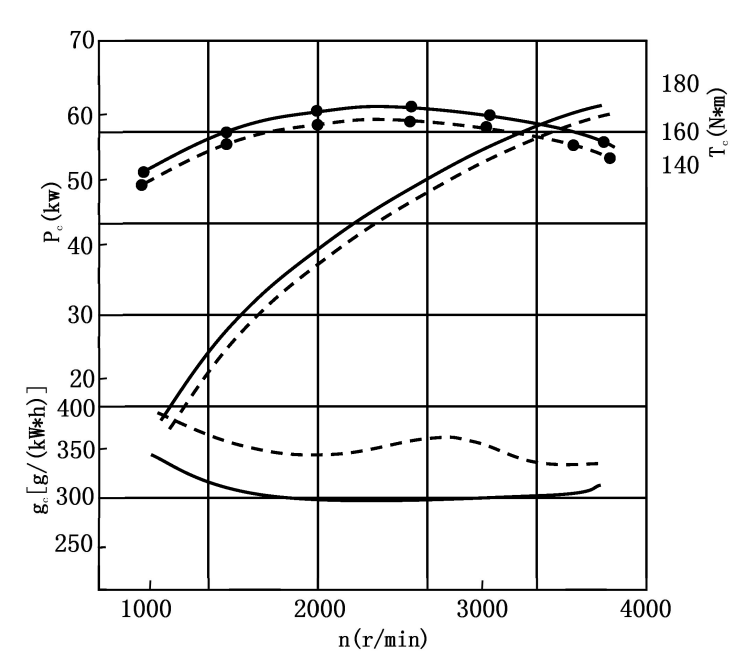} 
\caption{External characteristics of 492Q piston engine \cite{452000simulated}.}
\label{fig:11}
\end{figure}

\item[(iv)] Extend engine life

Aero-piston engines, owing to their complex structures and demanding operating conditions, are susceptible to various types of faults, with wear fault being the most common one. Engine tachometers often feature a red zone denoting speed limitations. While achieving these speeds is possible, prolonged operation within this range significantly accelerates engine wear. High engine speeds can lead to severe wear, but even at lower speeds, wear may not necessarily decrease. 

The engine's internal lubrication relies on an oil pump to deliver lubricant to critical areas, ensuring that the shaft is lifted with lubrication to prevent direct contact with bearings and reduce wear. The most substantial wear occurs when oil pressure is insufficient during startup. The oil pump is linked to the engine's crankshaft, and oil pressure increases with engine speed. Consequently, wear is exacerbated at low speeds and when the engine operates under heavy loads.

\item[(v)] Meet the special requirements

To comprehensively study the aerodynamic layout characteristics and flight dynamics of the original aircraft, it becomes imperative to establish specific similarity criteria between the original aircraft and the sub-scale test model. For propeller-driven aircraft, the preferred similarity principle often revolves around achieving similarity in the Froude number and the advance ratio \cite{462010Fundamentals}. 
The Froude number $\mathnormal{Fr}$ can be represented as:
\begin{equation} 
F r = \frac{V^{2}}{g l} \label{eq:1},
\end{equation}
where $\mathnormal{V}$ denotes the aircraft's velocity, $\mathnormal{g}$ stands for the gravitational coefficient, and $\mathnormal{l}$ represents the characteristic length of the aircraft. 

Given the equality of Froude numbers between the original aircraft ($\mathnormal{Fr_1}$) and the sub-scale test model ($\mathnormal{Fr_2}$), we can derive the following equations: 
\begin{equation} 
\label{eq:2}
\frac{V_1^{2}}{V_2^{2}} = \frac{l_1}{l_2} 
\end{equation}                                 
\begin{equation} 
\label{eq:3}
\frac{V_1}{V_2} = \sqrt{\frac{l_1}{l_2}}
\end{equation} 

The propeller advance ratio $\lambda$ can be represented as
\begin{equation} \label{eq:4}
\lambda = \frac{V}{n D},
\end{equation}
where $\mathnormal{n}$ denotes the RPM and $\mathnormal{D}$ stands for the diameter of the propeller. 
Assuming an equivalent advance ratio and based on the outcome of Eq. (\ref{eq:3}), we can derive 
 \begin{equation} \label{eq:6}
\frac{n_2}{n_1}= \frac{V_2 D_1}{V_1 D_2} = \frac{V_2 l_1}{V_1 l_2} = \sqrt{\frac{l_2}{l_1}} \cdot \frac{l_1}{l_2} = \sqrt{\frac{l_1}{l_2}}.
\end{equation} 
After establishing the scale of the aircraft, it becomes essential to maintain a constant rotational speed for both turboprop and turbo-shaft engines. Eq. (\ref{eq:6}) illustrates the specific ratio between these rotational speeds.

In the context of propeller power system similarity studies, it is often imperative to ensure similarity in Reynolds number and advance ratio \cite{liu2012study,462010Fundamentals}. The Reynolds number can be defined as
\begin{equation} \label{eq:7}
R e = \frac{\rho V l}{\mu},
\end{equation}   
where $\rho$ represents air density, $\mathnormal{V}$ denotes the aircraft velocity, $\mathnormal{l}$ stands for the characteristic length of the aircraft, and $\mu$ is the dynamic viscosity of air. 
Assuming equivalence in Reynolds number and advance ratio, we can derive 
 \begin{equation} \label{eq:8}
\frac{V_1}{V_2} = \frac{\rho_2 \mu_1 l_2}{\rho_1 \mu_2 l_1} = \frac{\rho_2 l_2}{\rho_1 l_1},
\end{equation} 
 \begin{equation} \label{eq:9}
\frac{n_2}{n_1} = \frac{V_2 D_1}{V_1 D_2} =  \frac{\rho_1 l_1}{\rho_2 l_2} \cdot \frac{D_1}{D_2} = \frac{\rho_1 l_1^{2}}{\rho_2 l_2^{2}}.
\end{equation} 
In many scenarios, the air density in the operational environment of the propeller prototype and the test air density of the sub-scale test model are determined. As indicated in Eq. (\ref{eq:9}), maintaining a constant rotational speed ratio between the scaled model and the prototype becomes imperative to fulfill the similarity requirements.

\end{enumerate}

\section{New challenges in constant speed variable-pitch control}

While the advantages of implementing effective variable pitch control at a constant rotational speed are evident, they also bring forth a set of new challenges. This approach allows us to harness the practical benefits mentioned earlier, whether individually or in combination. The dynamic interplay among the variable-pitch propeller, engine, and various associated mechanisms and accessories forms a highly complex propulsion system. Controlling such a system necessitates addressing a multifaceted array of factors including thermodynamics, aerodynamics, mechanics, electronics, vibration dynamics, and the influence of the atmospheric environment. 
Notably, this complexity introduces substantial nonlinear characteristics and uncertainties that manifest in various segments of the system and across different operational phases \cite{47Slotine2004Applied}. These challenges are persistent and omnipresent, stimulating ongoing research and the continuous development of pertinent engineering methodologies to overcome them.

\begin{figure*}[h]
\centering                                 
\includegraphics[width=\linewidth]{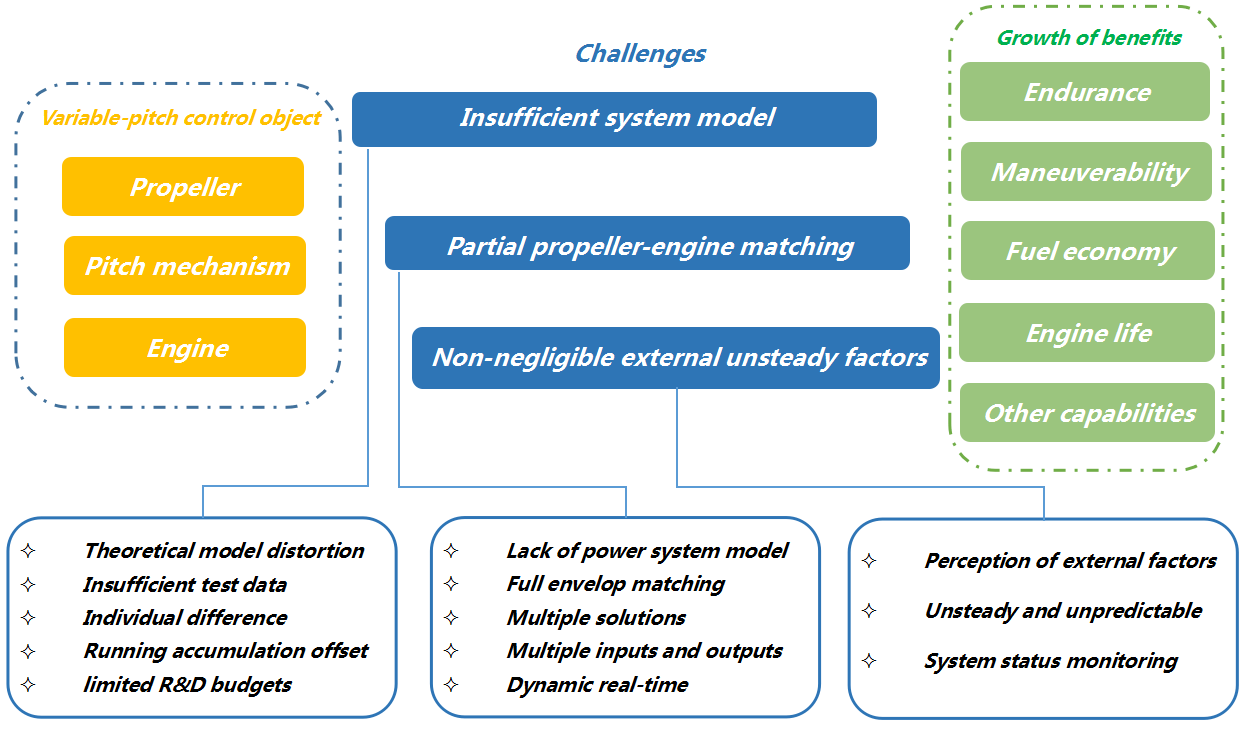} 
\caption{New challenges for constant speed variable-pitch control.}\label{fig:13}
\end{figure*}

Within this realm of technology, boasting a legacy spanning several decades, the focal point driving intense discourse on its challenges remains twofold: the pressing need for adaptability to new scenarios and heightened efficiency. Taking a closer look at the emerging challenges posed by the endeavor to achieve variable pitch control while maintaining a constant rotational speed, we can categorize them into three main domains: system modeling, engine-propeller compatibility, and external unsteady factors, as depicted in Figure \ref{fig:13}. Examining these fresh demands through the lens of engineering and specific scenarios, it becomes apparent that they hold the potential to usher in novel ideas and herald breakthroughs. This inherent capacity for innovation underlines the enduring value of these endeavors.

\subsection{Insufficient system model}

The system model primarily encompasses the engine and propeller models, with instances where the propeller is considered an integral part of the engine. Concerning the engine, model establishment generally adopts one of three approaches. 
The first method entails establishing a theoretical model for each engine component and, based on this foundation, formulating a mathematical engine model through the imposition of constraints and boundary conditions \cite{482004INTEGRATED}. 
The second approach relies on empirical data, with the engine's mathematical model derived through curve fitting of experimental, flight test, and calculated data \cite{492004Experimental,502015Full}. 
The third method combines the first two and is typically employed when developing an accurate theoretical model for certain engine components proves challenging \cite{512017Modeling}.

However, in numerous instances, obtaining precise theoretical and numerical engine models remains a formidable task. For example, the variable pitch control encountered in the sub-scale verification test aircraft discussed earlier is a typical technical challenge in the realm of constant speed variable pitch control. On one hand, the powerplants for simulated turboprop and turboshaft engine aircraft are usually small to medium-sized aviation piston engines or Wankel engines. For the majority of these engines, detailed test data to support an accurate numerical model is lacking. Furthermore, theoretical models often yield significant practical errors due to discrepancies between different engine types \cite{52El2016Fundamentals}. On the other hand, the use of small and medium-sized piston engines and rotor engines, which experience wear and exhibit issues like installation vibrations and exhaust pipe complications, introduces substantial variations in their operational characteristics compared to newly installed powerplants.

The challenge of inadequate system modeling is growing in scope, primarily due to the widespread adoption of small and medium-sized piston engines and rotor engines in UAVs and general aircraft. These aircraft have emerged as strong competitors across various industries, particularly in terms of operational costs. To establish a clear advantage over ground transportation vehicles and gain recognition and value within the logistics sector, unmanned aerial systems must significantly enhance their economic viability. 
One aspect of this challenge stems from budget constraints, preventing comprehensive testing similar to that performed on military aircraft engines, including ground and air-based tests. Additionally, to curtail operational costs, propulsion system efficiency and longevity are often achieved through fixed rotational speed variable pitch control. Consequently, addressing the model deficiency issue at a lower cost takes on broader significance. Such efforts are pivotal in providing the technical foundation required for constant speed variable pitch control, thereby exerting a profound impact on the application and advancement of the aviation industry.

\subsection{Partial propeller-engine matching}

Ensuring an appropriate relationship between the propeller and the engine is a critical aspect of enhancing overall aircraft performance and optimizing design. The process of engine-propeller matching typically relies on experimental testing methods \cite{53Lin2016Experimental} and theoretical design techniques \cite{542007Performance}. It centers on several key aspects, including aligning the power absorbed by the propeller with the engine's shaft power, matching propeller torque with engine shaft driving torque, and achieving rotational speed compatibility while adhering to limitations. These limitations primarily involve preventing the engine speed from falling below idle speed or exceeding its maximum speed \cite{woodward1973matching}.

In the case of larger power turboshaft and turboprop engines, ample test and design data are often available to support successful matching. However, for UAVs employing small and medium-sized piston engines, Wankel engines, general aviation aircraft, and vertical take-off and landing fixed-wing aircraft utilizing DC motors, the lack of power system models presents unique challenges to propeller-engine matching. Consequently, matching results achieved under such conditions are frequently limited in the scope of the full flight envelope, less efficient, and may even pose safety concerns.

Matching a fixed-pitch propeller with an engine typically revolves around meeting the primary requirement set by aircraft developers for the propulsion system: generating the necessary thrust at predetermined flight speeds. This type of matching centers on cruise point working conditions as the core, with other key flight conditions as secondary considerations. Such an approach simplifies the engine-propeller matching process significantly. 
Variable pitch propellers, on the other hand, allow for precise matching with the engine across most of the flight envelope range, thereby optimizing efficiency. 
However, if matching constraints are solely based on thrust requirements at predetermined flight speeds, the challenge of multiple balance points emerges. Introducing the additional requirement of fixed rotational speed resolves the issue of multiple balance points while delivering additional performance benefits. However, this places higher demands on the precision of propeller-engine matching, particularly when addressing matching and control under dynamic flight conditions.

\subsection{Non-negligible external unsteady factors}
During flight through the atmosphere, an aircraft's dynamic system's control and operational state are inevitably influenced by external factors, particularly when these factors are time-varying or sudden changes. Many high-speed and large-scale aircraft often simplify the impact of external unsteady factors, treating them as minor disturbances. However, for low-speed and small to medium-sized aircraft, these external environmental disturbances cannot be dismissed. 

\begin{figure}[ht]
\centering                             
\includegraphics[width=0.8\linewidth]{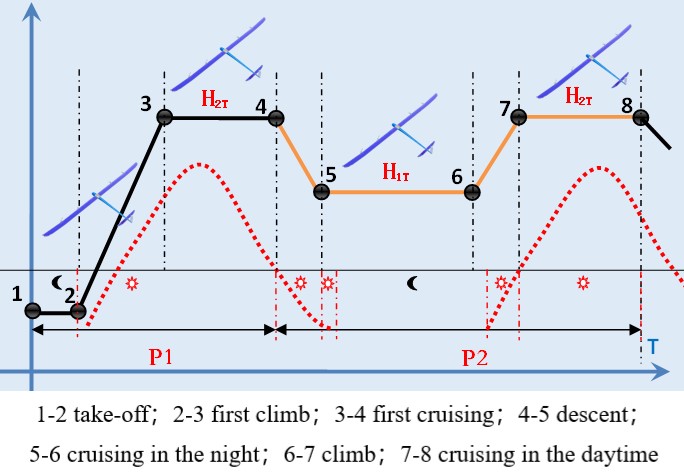} 
\caption{Common flight profile of solar powered HASLE UAV \cite{Padilla2020Flight}}.
\label{fig:12}
\end{figure}

Taking these disturbances seriously demands significant efforts, often of a comprehensive nature. For example, High Altitude Super-Long Endurance (HASLE) solar-powered UAVs prioritize cruise efficiency and aim for increasingly extended flight durations. The random meteorological factors these UAVs may encounter during flight can profoundly affect efficiency goals and even UAV safety. 
Traditionally, UAVs operated on predefined flight profiles crossing day and night, as illustrated in Figure \ref{fig:12}. These profiles were based on theoretical considerations and flight test experience \cite{55Oh20163}, and included the corresponding implementation of propeller variable pitch control. Looking ahead, high-altitude, long-endurance solar-powered UAVs will necessitate online learning to construct variable flight profiles and corresponding pitch angle control strategies to adapt to evolving environmental conditions in real time.

The principal challenges revolve around controller design, enhancing the accuracy of engine and propeller models, real-time online control optimization, and the verification of control system tests. Initial steps involve shedding fixed power unit digital models to minimize model inaccuracies, regardless of whether they are derived from experiments or theory. Thorough consideration must be given to random environmental factors and the cumulative differences that emerge as the aircraft is in use. Simultaneously, there is a growing expectation for UAVs to exhibit increasing intelligence, demanding that the aircraft's control system be autonomous, adaptive, and able to learn online and in real-time.

\section{Conclusion}

The history of aircraft development spans over a century, yet propeller-driven aircraft remain prevalent and vital in aviation. Pitch control technology, as a paramount means to enhance the efficiency of propeller power systems, finds extensive application in helicopters, propeller-driven transporters, and ultralight aircraft. And more recently, it also plays an important role in the burgeoning realm of multi-rotor drones, fixed-wing unmanned aircraft, and urban air mobility concepts. The continuous advancement of aircraft technology has elevated the demands placed on this traditional but critical technology.

Variable-pitch technology, with its diverse mechanisms and principles, empowers aircraft and pilots to achieve desired outcomes such as thrust, pull, acceleration, rotational speed, or optimal dynamic efficiency while ensuring stable flight. The effective coordination of propellers and powerplants, with precise control systems, is pivotal in realizing these expectations.

In the pursuit of these goals, certain challenges have come to the forefront:
\begin{enumerate}
\item[(i)] The inaccuracy or even the absence of a dynamic model.
\item[(ii)] Considerable variations in engine states, which escalate with engine wear.
\item[(iii)] The reluctance or inability of pilots to participate fully in flight, while power control laws lack specificity in certain aspects.
\end{enumerate}
It is evident that the advancement of intelligent technologies, such as machine learning and depth perception, will play a pivotal role in addressing these challenges and steering the future of aviation towards greater efficiency and precision.

\section*{Acknowledgments}
This work is sponsored by Universiti Sains Malaysia (USM) with the Short Term Research Grant Scheme [grant number 304/PAERO/6315297].

\bibliographystyle{unsrt}  
\bibliography{Reference}

\end{document}